\documentclass[aps,pra,groupedaddress,twocolumn,showpacs,showkeys]{revtex4}
\usepackage{graphicx}
\usepackage{amssymb}
\usepackage{subfigure}
\usepackage{amsmath}
\usepackage{float}
\usepackage{lipsum}
\usepackage{braket}
\usepackage{amsfonts}
\usepackage{bm}
\usepackage{color}
\usepackage{dsfont}
\newcommand{\beq}{\begin{equation}}
\newcommand{\eeq}{\end{equation}}

\begin{document}

	\preprint{}

	\title{3D compact photonic circuits for realizing quantum state tomography of qudits in any finite dimension} 
	\author{W. R. Cardoso}
	\email{will.rodrigues.fis@gmail.com}
	\author{D. F. Barros}
	\author{L. Neves}
	\author{S. P\'adua}
	\affiliation{Departamento de F\'{\i}sica, Universidade Federal de Minas Gerais, 31270-901 Belo Horizonte, Minas Gerais, Brazil}

	\begin{abstract}
    
    In this work, we propose three-dimensional photonic circuit designs that guarantee a considerable reduction in the complexity of circuits for the purpose of performing quantum state tomography of $N$-dimensional path qudits. The POVM (Positive Operator-Valued Measure) chosen in this work ensures that, for odd dimensions, such process is minimal. Our proposal consists of organizing the waveguides that form the circuit as a square array formed by $N$ vertical sectors composed of $N$ waveguides each, arranged in the vertical direction. Based on the symmetry of the chosen POVM, the interferometer acting on the initial quantum system can be divided into a sequence of three different unitary operations. These operations act independently on each vertical sector, or layer, of the circuit, which simplifies their determination. We have thus obtained circuits such that the number of beam splitters obeys a polynomial function of degree 3 with the quantum system dimension, whereas in current proposals this quantity grows with a polynomial function of degree 4. Besides that, the optical depth is reduced from a quadratic to a linear function of the quantum system dimension in our scheme. These results confirm the remarkable reduction of the complexity of the photonic circuits in our proposal.    
    
	\end{abstract}


	\maketitle
	\section{Introduction}
	\label{sec:intro}
	
	An important experimental achievement within the field of quantum information is the ability to reconstruct density matrices of unknown quantum states. This is done through the technique of quantum state tomography \cite{pimenta2010minimal,pimenta2013minimum,titchener2018scalable,toninelli2019concepts,olivares2019quantum,titchener2016two}. Such technique consists in the realization of a specific set of measurements on several identically prepared quantum states and, from the results of these measurements, reconstruct its density matrix \cite{bayraktar2016quantum,munro2001measurement,ling2006experimental}.	Quantum state tomography has been used to determine the density matrix of many different quantum systems, such as the spin and the energy levels of trapped ions \cite{home2006deterministic,riebe2006process}, molecules \cite{nielsen1998complete}, and photons \cite{munro2001measurement,pimenta2013minimum,ansari2018tomography,de2019overcomplete,toninelli2019concepts,quijandria2018quantum}.
	
	There are many ways to perform quantum state tomography \cite{gross2010quantum,kyrillidis2018provable,bohm2018pump,fernandez2011quantum,lima2011experimental,klimov2008optimal}, but, in this work, we choose the one based on POVMs via equidistant states \cite{paiva2010quantum}. This approach ensures that the unknown state of the quantum system is reconstructed with the minimum number of measurement outcomes, without any previous information about it. This feature guarantees a minimization of the experimental errors involved in the process. An $N$-dimensional POVM that implements minimum quantum tomography has $N^2$ elements. It is shown in Ref. \cite{paiva2010quantum} that a POVM tomography via equidistant states is able to reconstruct an $N$-dimensional density matrix with $N^2$ measurements when $N$ is odd and $3N^2/2$ measurements when $N$ is even. This larger number of measurements in the case of even dimension, in relation to the odd dimension case, is justified by a small modification made in the initial state before passing through the measurement apparatus, a necessary extra step to avoid the cancellation of some terms of the density matrix in the Born rule, preventing its determination. 
	
	Knowing that each quantum operation is done experimentally through a specific setup on an optical table, performing a set of operations requires several reconfigurations of such setup, increasing the errors associated with the measurements and the time spent in the experiment. An alternative to all this is the use of static photonic circuits \cite{oren2017quantum}. Static circuits are those where the quantities that define it are kept fixed during the measurement process. Thus, it is possible to use a single apparatus where all measurements are made simultaneously to perform quantum state tomography optimally. This type of circuit, besides being stable and scalable \cite{sansoni2014integrated}, presents itself as more compact than others, since it can be constructed using single beam splitters as building blocks, instead of using Mach-Zehnder interferometers \cite{clements2016optimal}. Photons at the exit of the quantum tomographic circuit can be detected by click detectors, which simplifies the setup. Photon number resolving detectors are not necessary for this application \cite{heilmann2016harnessing,luis2015nonclassicality,hlouvsek2019accurate}.
	
	In this work, we seek to develop a method of obtaining the design of a 3D photonic circuit that performs full quantum state tomography in one-qudit systems of any finite dimension $N\geqslant3$ represented, in practice, in the base of photon-path states. The photon-path states in a photonic circuit are defined by $N$ waveguides and form a state space of dimension $N$. The qudit input state, for one input single photon, is a state written in terms of these $N$ photon-path states. Our method has no advantage in the case of qubits in relation to planar circuits \cite{tabia2012experimental,wilder}. 
	
	The paper is divided in three sections. In Section II, we define the equidistant states, construct through them the POVM used in the full quantum state tomography and relate the probabilities associated with the POVM elements with the density matrix elements of the quantum state under study. In Section III, our proposal itself is presented. We use the strategy of dividing the interferometer into three distinct unitary operations that act independently between layers or vertical sectors. As a result, we have achieved a considerable reduction in the complexity of the photonic circuit. The number of beam splitters in our proposal scales as a lower degree polynomial function than in other known proposals \cite{clements2016optimal,reck1994experimental}. The same happens with the optical depth, quantity defined as the maximum number of beam splitters traversed by a photon from its input port until its output port from the circuit. Our conclusions are presented in Section IV.	
	
	\section{Equidistant States and POVM elements}
	\label{sec:equidistant}

   	For the construction of the POVM used in our proposal of photonic circuits for quantum state tomography, we are interested in a set of $N$ nonorthogonal state vectors $A_N(\ket{\varphi_m})$ that satisfy the following condition
   	\begin{equation}
   		\braket{\varphi_m|\varphi_n}=\alpha=|\alpha|e^{i\theta}, \qquad\quad \forall m\neq n.
   		\label{iner}
   	\end{equation}
   
   	The inner product between any two elements of $A_N(\ket{\varphi_m})$ is always equal to $\alpha$ or its complex conjugate $\alpha^*$. Due to this property, the vectors of this set are called equidistant states \cite{paiva2010quantum,jimenez2010probabilistic}. The analysis of the Gram determinant makes it possible to determine if $A_N(\ket{\varphi_m})$ is linearly independent (LI) or linearly dependent (LD) \cite{roa2011conclusive}. This determinant is defined as 
   	\begin{equation}
   		D_{N\times N}=\text{det}
   		\left[\begin{array}{ccccc}
   		1 & \alpha & \alpha & \cdots & \alpha \\
   		\alpha^* & 1 & \alpha & \cdots & \alpha \\
   		\alpha^* &  \alpha^* & 1 & \cdots & \alpha \\
   		\vdots & \vdots & \vdots & \ddots & \vdots \\
   		\alpha^* &  \alpha^* & \alpha^* & \cdots & 1 \\   		
   		\end{array}\right]_{N\times N},
   	\end{equation}
   	which can be rewritten as
   	\begin{equation}
   		D_{N\times N}=\dfrac{\alpha\left(1-\alpha^*\right)^N-\alpha^*\left(1-\alpha\right)^N}{\alpha-\alpha^*}.
   	\end{equation}
   	
   	The set $A_N(\ket{\varphi_m})$ is linearly independent if, and only if, $D_{N\times N}\neq0$. Therefore, as already discussed in previous works \cite{roa2008petal,roa2011conclusive}, in a LI set of vectors, the absolute values of the inner product $\alpha$ must satisfy the following restriction  
   	\begin{equation}
   		0\leqslant |\alpha| < |\alpha_\theta^{\text{LD}}|,
   	\end{equation}
   	where $\alpha_\theta^{\text{LD}}$ is the value of the inner product when such a set is LD. The constant $\alpha_\theta^{\text{LD}}$ depends directly on the quantities $\theta$ and $N$ as follows \cite{paiva2010quantum,roa2011conclusive}
   	\begin{equation}
   		\alpha_\theta^{\text{LD}}=\text{\scriptsize$\dfrac{\sin\left(\dfrac{\pi-\theta}{N}\right)}{\sin\left(\theta+\dfrac{\pi-\theta}{N}\right)}$}, \qquad\quad 0\leqslant \theta <2\pi.
   	\end{equation}
   	
   	Another known result is that a set of equidistant states is also symmetric when the inner product $\alpha$ is a real number, that is, $\theta=0$ or $\theta=\pi$ \cite{paiva2010quantum,roa2008petal}. In these cases, we obtain $\alpha_0^{\text{LD}}=1$ and $\alpha_\pi^{\text{LD}}=(N-1)^{-1}$, respectively.
   	
   	The canonical decomposition of the equidistant states is given by \cite{paiva2010quantum,jimenez2010probabilistic} 
   	\begin{equation}
   	\ket{\varphi_{m}}=\dfrac{1}{\sqrt{N}}\sum_{k=0}^{N-1}\left(\omega_k\right)^m\sqrt{\lambda_k}\ket{k}, 
   	\end{equation}
   	where the state vectors set $\{\ket{k}\}$ represents an orthonormal basis in the Hilbert space, $m=0,1,...,N-1$,
   	\begin{equation}
   	\omega_k=e^{-2i\left(\theta-k\pi\right)/N},
   	\label{omega}
   	\end{equation}
   	and
   	\begin{equation}
   		\lambda_k=1-|\alpha|\text{\scriptsize$\dfrac{\sin\left(\theta+\dfrac{k\pi-\theta}{N}\right)}{\sin\left(\dfrac{k\pi-\theta}{N}\right)}$}.
   		\label{lambda}
   	\end{equation}
   	
   	Thus, to ensure symmetry properties, $\theta=\pi$ is chosen, since $\theta=0$ generates null state vectors. With this value for $\theta$, the canonical decomposition is simplified to
   	
   	\begin{equation}
   	\ket{\varphi_m}=\dfrac{1}{\sqrt{N-1}}\sum_{\substack{k=0 \\ k\neq 1}}^{N-1}e^{2i\pi m\left(k-1\right)/N}\ket{k}.
   	\label{state}
   	\end{equation}  	
   	
   	The index $m$ assumes $N$ different values, generating a number of state vectors smaller than necessary to construct all the POVM elements required by the process of quantum state tomography. Thus, it is necessary to derive other sets of equidistant states from the initial set $A_N(\ket{\varphi_m})$ by the application of a specific operator $\hat{X}$ which acts as follows \cite{paiva2010quantum}
   	\begin{equation}
   		\hat{X}\ket{k}=\ket{k\oplus 1}, 
   	\end{equation}
   	where the symbol `$\oplus$' represents an addition modulo $N$. These new vector sets are defined as
   	\begin{equation}
   		A_N^{(l)}(\ket{\varphi_m})=\{\ket{\varphi_{ml}}=\hat{X}^l\ket{\varphi_m}\},
   		\label{anl}
   	\end{equation}   	
   	where $l=0,1,...,N-1$. The POVM elements needed for the tomographical process are constructed from the states $\ket{\varphi_{ml}}$ of Eq.~\eqref{anl} by the relation
   	\begin{equation}
   	\hat{E}_{ml}=\dfrac{1}{N}\ket{\varphi_{ml}}\bra{\varphi_{ml}}.
   	\label{eij}
   	\end{equation}
   	
   	The constant $1/N$ comes from the completeness property of the POVM \cite{barnett2009quantum}. The POVM formed by the elements $\hat{E}_{ml}$ defined in Eq.~\eqref{eij} is a SIC-POVM (Symmetric Informationally Complete POVM) only for the particular case where $N=3$ \cite{paiva2010quantum}. The calculation of the probability associated with each POVM element is done by the Born statistical formula $P_{ml}=\text{Tr}(\hat{\rho}\hat{E}_{ml})$, resulting in
   	\begin{equation}
   	P_{ml}=\dfrac{1}{N}\sum_{r,s=0}^{N-1}\rho_{rs}e^{2i\pi m(r-s)/N}\sqrt{\lambda_{r-l}\lambda_{s-l}},
   	\label{prob}
   	\end{equation}
   	where $\hat{\rho}$ is the density operator that represents the unknown state in an $N$-dimensional Hilbert space and the subtractions in the subindexes of $\lambda$ are carried out modulo $N$. Thus, in possession of all experimental probabilities $P_{ml}$, it is possible to obtain a $N^2$ equations system with $N^2$ variables through the Eq.~\eqref{prob}. Solving them, one obtains the values of all matrix elements $\rho_{rs}$, and consequently, the density matrix of the system is determined.
   	
   	A setback was reported in the even-dimension cases \cite{paiva2010quantum}. Such a problem is perceived most clearly when we rewrite the equation that defines the probability $P_{ml}$ in a different way. When rewriting the Eq.~\eqref{prob}, we see that, in its new form, the following term appears   	
   	\begin{equation}
   		\sum_{r>s}\left[\text{Re}(\rho_{rs})\cos\Omega-\text{Im}(\rho_{rs})\sin\Omega\right]\sqrt{\lambda_{r-l}\lambda_{s-l}},
   	\end{equation}
   	where $\Omega=2\pi m(r-s)/N$. When $N$ is even, we will have, in some cases, $r-s=N/2$, which implies $\Omega=m\pi$ and, thereby, $\sin\Omega=0$. In such cases, the determination of $\text{Im}(\rho_{rs})$ becomes impossible. To overcome this problem, a quantum operation is applied to the density matrix of the system in order to enable the determination of such density matrix elements. The price paid for performing this extra step in the process is to increase the number of measurements required from $N^2$ to $3N^2/2$ \cite{paiva2010quantum}.  	
	\section{Photonic Circuits Design}
	\label{sec:design}

	In our previous work \cite{wilder}, we discussed the tomographical photonic circuits for qubit and qutrit quantum systems. Planar circuits were proposed for these two cases and, considering the practical inviability of designing a planar circuit for the case $N=4$, we made brief comments on how interesting is the use of three-dimensional architectures for manufacturing photonic circuits for quantum tomography. The generalization of tomographical photonic circuits for $N$-dimensional states $(N\geqslant 3)$ are discussed here in detail. 
	
\begin{figure}[]
		\centering
		\includegraphics[height=5.5cm]{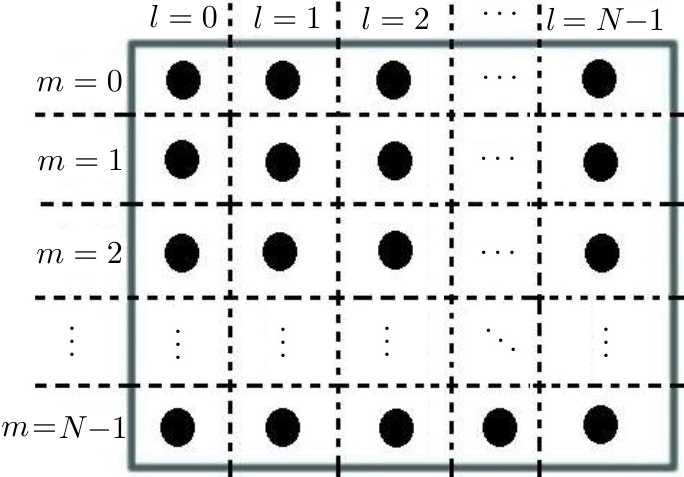}
		\caption{Transverse section of the photonic circuit inputs. The black circles represent the waveguides. The vertical sectors are labeled by the letter $l$ and the layers, by the letter $m$.}
		\label{frente}
	\end{figure}

	We use a 3D structure for the circuits since this choice allows us to design more compact layouts. Our interferometer is constructed as a square array of $N^2$ waveguides, where $N$ parallel vertical sectors contains $N$ waveguides arranged in the vertical direction. The transverse section of the circuit is shown in Fig.~\ref{frente}, where the vertical sectors are labeled by the letter $l$ and the layers, by the letter $m$. Thus, each path state will be identified as a vector $\ket{ml}$ $(m,l=0,1,...,N-1)$. In this proposal, the path qudits must enter the circuit through the inputs belonging to the layer $m=0$ or layer $m=N-1$. 
	
	It was already proved that the used equidistant states POVM is informationally complete \cite{paiva2010quantum}. Therefore, this POVM is able to realize quantum tomography of any quantum general state, including general mixed state. Therefore, a photonic circuit that implements this equidistant states POVM is able to realize quantum tomography in any input qudit general mixed state. We will use a input pure state in the circuit derivation only because is more pedagogical, but our proposal is general for any input qudit state quantum tomography. So, we chose as input state	
	\begin{equation}
		\ket{\psi}=\sum_{l=0}^{N-1}a_l\ket{0l}.
	\end{equation}
	
	The proposed photonic circuit is divided into three parts, each one implementing a different unitary operation. These operations act sequentially, that is, the final state in one of them serves as the initial state for the next one. These operations will be discussed in detail in the following subsections.	
	
	\subsection{First Part: Decomposition}
	
		\begin{figure}[]
		\centering
		\includegraphics[height=4.2cm]{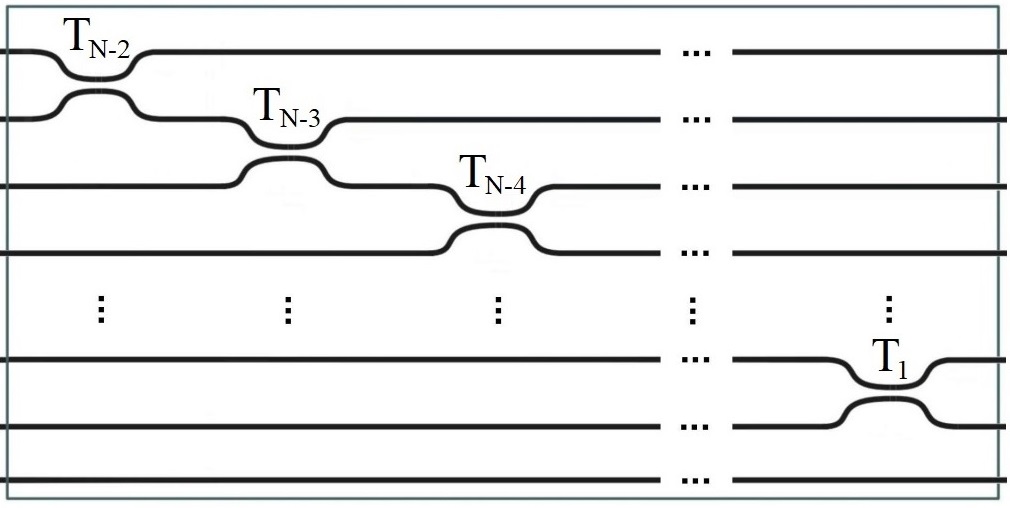}
		\caption{Arrangement of beam splitters for the first part of the tomographical photonic circuit. The black lines are the waveguides. The beam splitters, labeled by $T_k$ ($k=1,2,...,N-2$), are represented by the approximation of these lines.}
		\label{decomp}
	\end{figure}
	
	In the first part of the circuit, we are interested in a unitary operation that acts in each vertical sector $l$ independently and realize the following transformation on each component of the input state 
	\begin{equation}
		a_l\ket{0l} \quad \longmapsto\quad \dfrac{a_l}{\sqrt{N-1}}\sum_{m=0}^{N-2}\ket{ml}.
		\label{trans1}
	\end{equation}
	
	Thus, it creates a superposition of $N-1$ components, occupying almost all the paths of each vertical sector. The arrangement of beam splitters that realize this transformation is shown in Fig.~\ref{decomp} and is formed by a diagonal line composed of $N-2$ beam splitters. The transmittance $t_k$ and the reflectivity $r_k$ of the beam splitter $\hat{T}_{k}$, where $k=1,2,...,N-2$, in Fig.~\ref{decomp}, are defined by
	\begin{equation}
	t_k=\dfrac{k}{k+1}, \qquad\qquad  r_k=\dfrac{1}{k+1},
	\end{equation}
	
	The same arrangement will appear in all vertical sectors, so that $N(N-2)$ beam splitters will be needed to implement this first part of the circuit. This part has an optical depth of $N-2$.
	
\begin{figure}[]
		\centering
		\includegraphics[height=3.15cm]{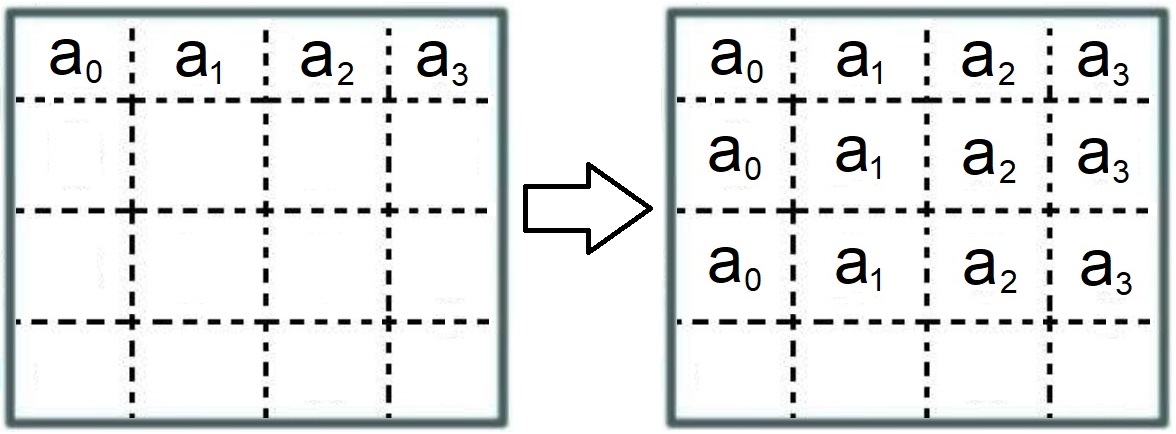}
		\caption{Schematic representation of the decomposition operation. On the left, we have the distribution of the coefficients of the ququart state $\sum_{l=0}^{3}a_l\ket{0l}$ at the input ports of the first part of the circuit. Here, only the first layer paths are occupied. On the right, we present the distribution of the state coefficients after the application of the decomposition operation. The coefficients $a_l$ were diffused by $N-1$ layers in each vertical sector $l$. The normalization term $1/\sqrt{3}$ was omitted in the figure on the right.}
		\label{u1}
	\end{figure}

	In order to elucidate the application of this first transformation, consider as an example a ququart system ($N=4$). Assuming that the input state is $\sum_{l=0}^{3}a_l\ket{0l}$, Fig.~\ref{u1} presents the schematic distribution of its coefficients before and after the application of the decomposition operation performed in this first part of the circuit.
		
	\subsection{Second Part: Permutation}	
	
	\begin{figure}[]	
		\center
		\subfigure[ref1][For $m=N-4$ and $\upsilon=3$]{\includegraphics[width=8.5cm]{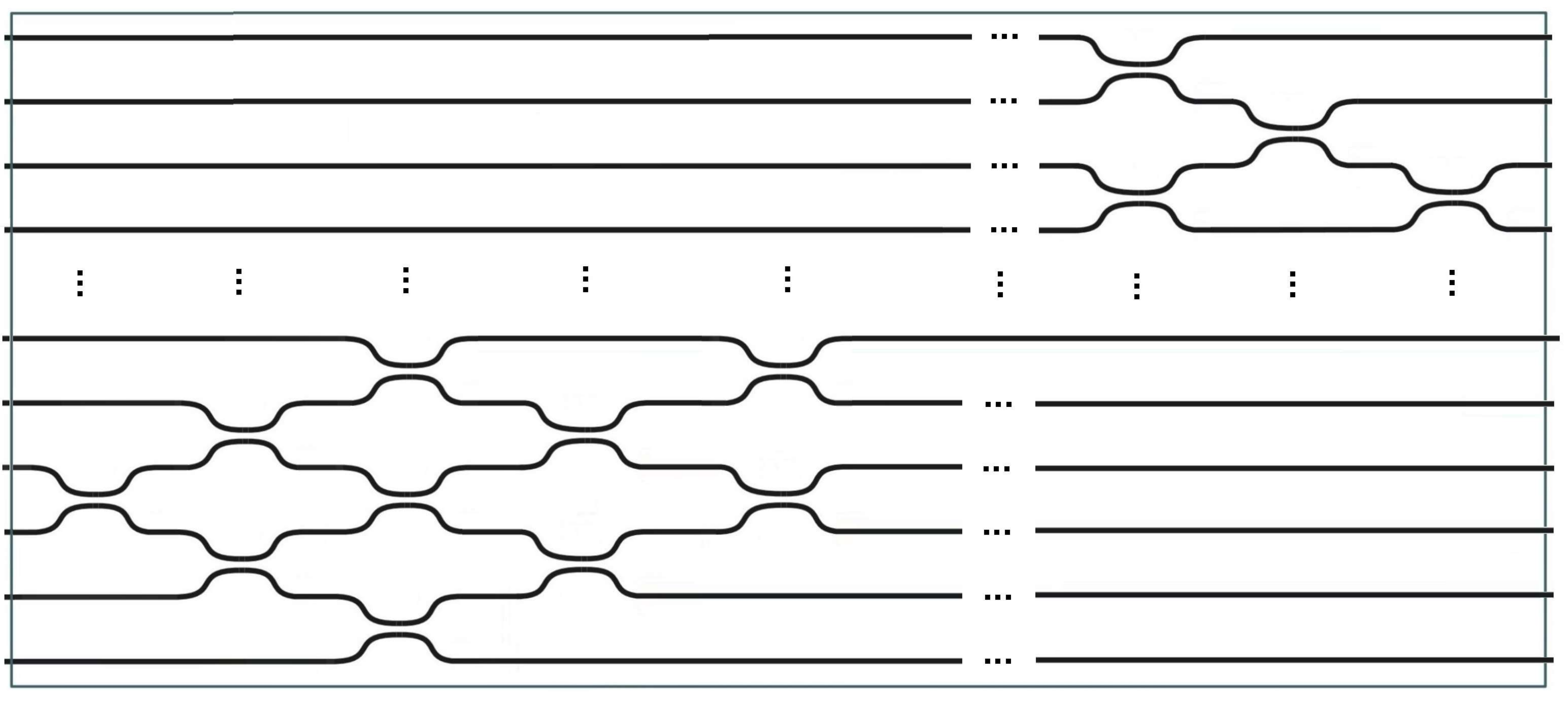}}
		\subfigure[ref2][For $m=N-3$ and $\upsilon=2$]{\includegraphics[width=8.5cm]{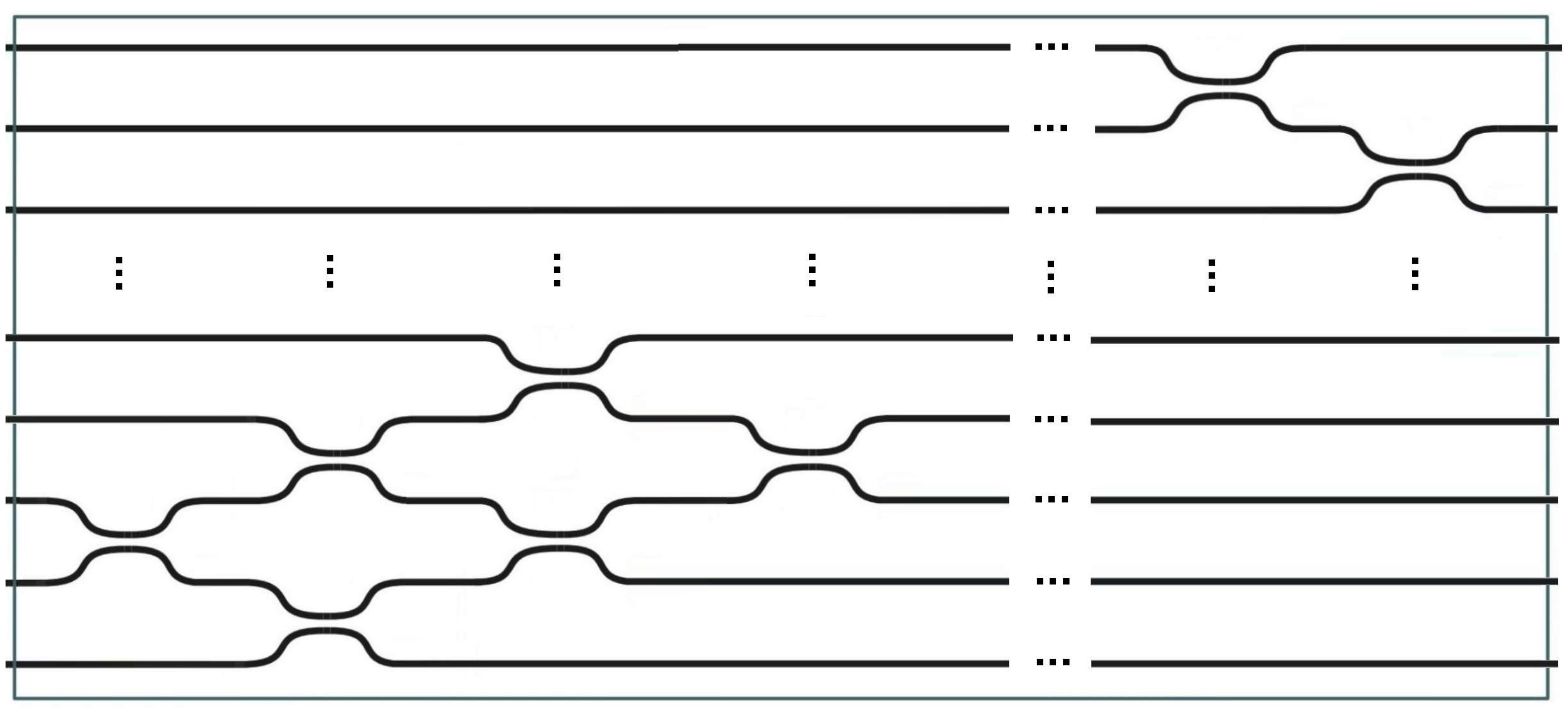}}
		\subfigure[ref3][For $m=N-2$ and $\upsilon=1$]{\includegraphics[width=8.5cm]{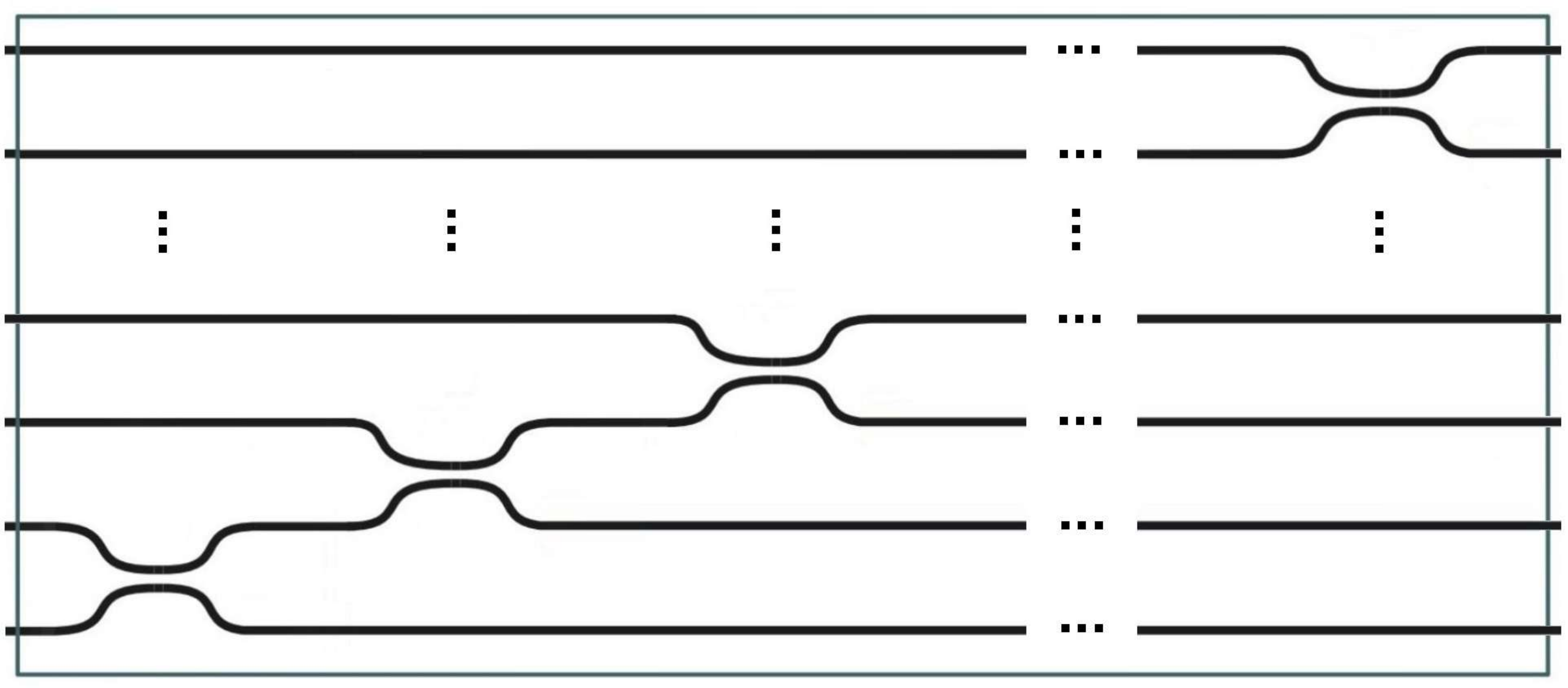}}
		\caption{Beam splitters arrangement that realizes the permutation $\hat{\varGamma}_m$ in the last layers of the interferometer, for the cases where $m\geqslant 1$. All beam splitters have transmittance equal to 1.}	
		\label{permut}
	\end{figure}
	
	\begin{figure}[]
		\centering
		\includegraphics[height=3.15cm]{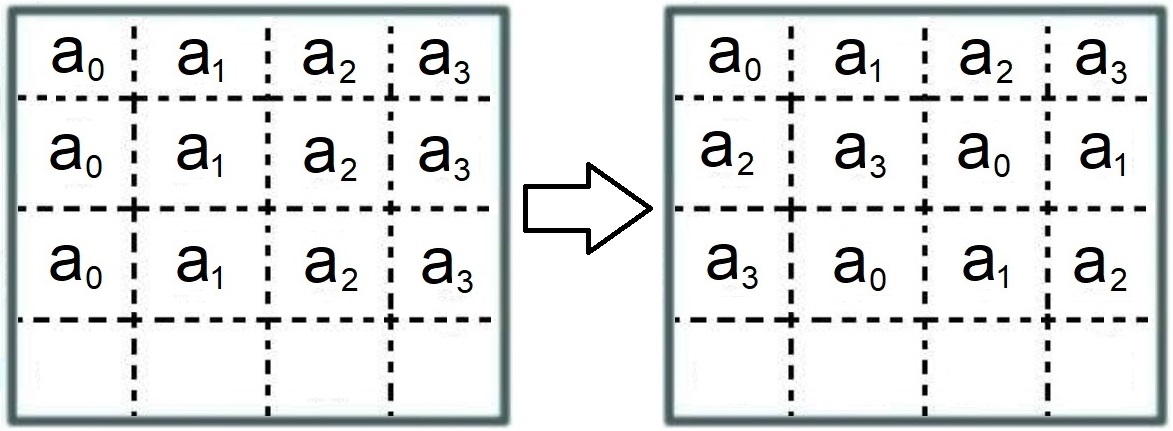}
		\caption{Schematic representation of the permutation operation. The left and right frames are, respectively, the input and output of the circuit second part. On the right, in each vertical sector, we have the coefficients of the base states which  form the equidistant states in Eq.~\eqref{anl}. The normalization term $1/\sqrt{3}$ was omitted.}
		\label{u2}
	\end{figure}		
	
	Unlike the first part of the circuit, the permutation operation in the second part of the interferometer acts independently in each horizontal layer, between vertical sectors. The operation made on the input state in this part of the circuit is different in each layer and can be represented by the following relation
	\begin{equation}
		\hat{\varGamma}_m\ket{ml}=\ket{m}\ket{l\oplus\upsilon},
		\label{gama}
	\end{equation}
	where $\upsilon=N-(m+1)$ and $m\geqslant 1$. An addition modulo $N$ is realized in the sector index $l$ with such operation being a function of the layer index $m$. Thus, in each vertical sector $l$, the following transformation is performed 
	\begin{equation}
		\dfrac{1}{\sqrt{N-1}}\sum_{m=0}^{N-2}a_l\ket{ml} \longmapsto \dfrac{1}{\sqrt{N-1}}\sum_{m=0}^{N-2}a_{l-\upsilon}\ket{ml},
		\label{eq2}
	\end{equation}
	where the subtraction in the subindex of $a_{l-\upsilon}$ is carried out modulo $N$. The last three permutations of a general circuit are shown in Fig.~\ref{permut}. Analyzing the layers it is possible to see a pattern in the arrangement of beam splitters. The arrangement in the layer $m$ is formed by $\upsilon$ diagonal lines with $N-\upsilon$ beam splitters each. Thus, the number of beam splitters needed to implement this second part of the interferometer is equal to $\sum_{\upsilon=1}^{N-2}\upsilon(N-\upsilon)$ and it has an optical depth of $N-1$. Another interesting feature of this part of the circuit is that all beam splitters are equal. All of them only transmit, that is, they have a unit transmittance ($t=1$) and a null reflectivity ($r=0$).
	
	Following our example from the previous subsection, a ququart state, we present in Fig.~\ref{u2} the action of the second part of the circuit, explaining the distribution of its coefficients by the interferometer sites before and after the application of the permutation operators $\hat{\varGamma}_m$, defined in Eq.~\eqref{gama}.

	\subsection{Third Part: Fourier Transform}
	
	The third and last part of the circuit is responsible for the execution of a Fourier transform. This transformation is done independently in each vertical sector. It acts on the post-processing quantum state given by Eq.~\eqref{eq2} and transforms it into another quantum state of the form $\sum_{ml}\beta_{ml}\ket{ml}$. This transformation is the same for all vertical sectors and is represented by a unitary matrix $\hat{U}$ that performs the following operation
	\begin{equation}
		\hat{U}\cdot\begin{pmatrix}
		a_{l+1-N}/\sqrt{N-1}\\a_{l+2-N}/\sqrt{N-1}\\\vdots\\a_{l-1}/\sqrt{N-1}\\0
		\end{pmatrix}=\dfrac{1}{\sqrt{N(N-1)}}\begin{pmatrix}
		\beta_{0l}\\\beta_{1l}\\\vdots\\\beta_{(N-2)l}\\\beta_{(N-1)l}
		\end{pmatrix},
	\end{equation}
	where the operations in the subindexes of $a_{l-\upsilon}$ are carried out modulo $N$ and the coefficients $\beta_{ml}$ are defined as
	\begin{equation}
		\beta_{ml}=\sum_{\substack{k=0 \\ k\neq 1}}^{N-1}a_{k+l}e^{2i\pi m\left(k-1\right)/N},
		\label{beta}
	\end{equation}
	where, again, the operations in the subindexes of $a_{k+l}$ are carried out modulo $N$. Equation \eqref{beta} guarantees the validity of the following relation
	\begin{equation}
	\dfrac{1}{N(N-1)}|\beta_{ml}|^2=\text{Tr}\left(\hat{\rho}\hat{E}_{ml}\right),
	\end{equation}
	i.e., the probability of detecting a photon emerging from one of the circuit outputs is equal to the probability associated with the POVM element $\hat{E}_{ml}$, defined in Eq.~\eqref{eij}. When we obtain these probabilities experimentally, we are able to reconstruct the density matrix of the system under study with the use of Eq.~\eqref{prob}.
		
	To ensure that the last part of the circuit perform such a transformation, it is necessary to implement the following unitary operation described by the matrix \cite{lim2005multiphoton}
	\begin{equation}
		\hat{U}=\dfrac{1}{\sqrt{N}}\begin{pmatrix}
		1 & 1 &  \cdots & 1 & 1 \\
		\sigma & \sigma^2 &  \cdots &   \sigma^{N-1} & 1 \\
		\sigma^2 & \sigma^4 &  \cdots &  \sigma^{2(N-1)} & 1 \\
		\vdots & \vdots &  \ddots & \vdots & \vdots \\ 
		\sigma^{N-1} & \sigma^{2(N-1)} &  \cdots &  \sigma^{(N-1)(N-1)} & 1
		\end{pmatrix},
		\label{um}
	\end{equation}
	where $\sigma=e^{2i\pi/N}$. The implementation of this last part is done using an analogous scheme to the Clements's one \cite{clements2016optimal}, where the Mach-Zehnder interferometers are replaced by single beam splitters. This exchange is justified by the fact that our proposal consists of static circuits.
	
	\begin{figure}[]	
		\center
		\subfigure[ref1][$N=3$]{\includegraphics[width=8.5cm]{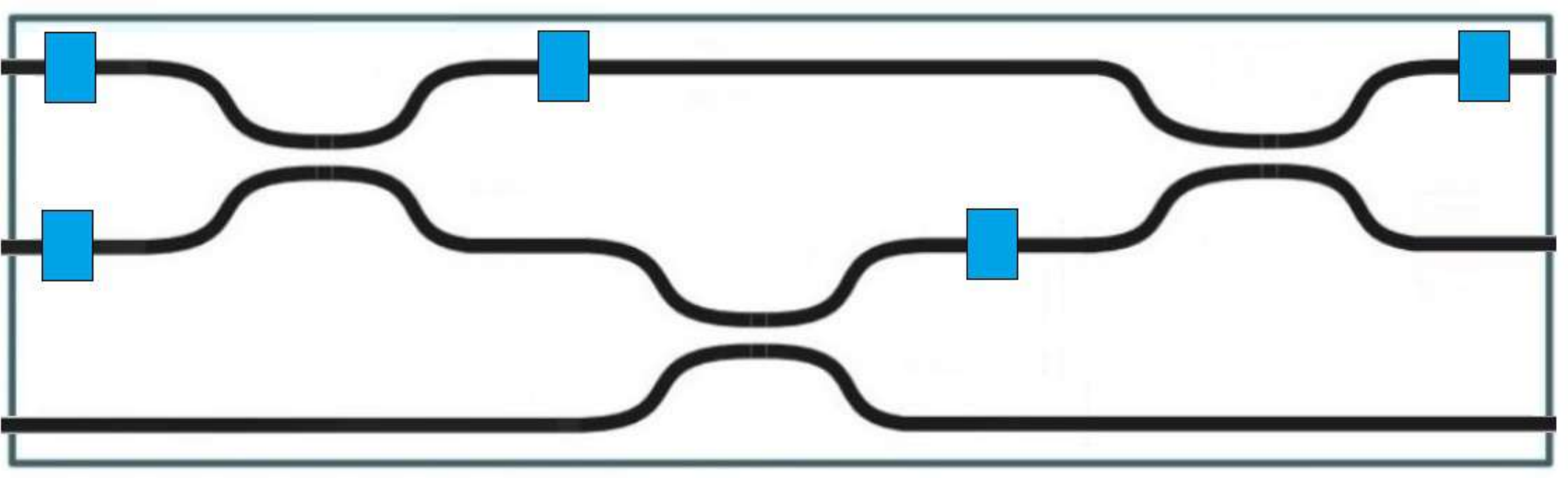}}
		\subfigure[ref2][$N=4$]{\includegraphics[width=8.5cm]{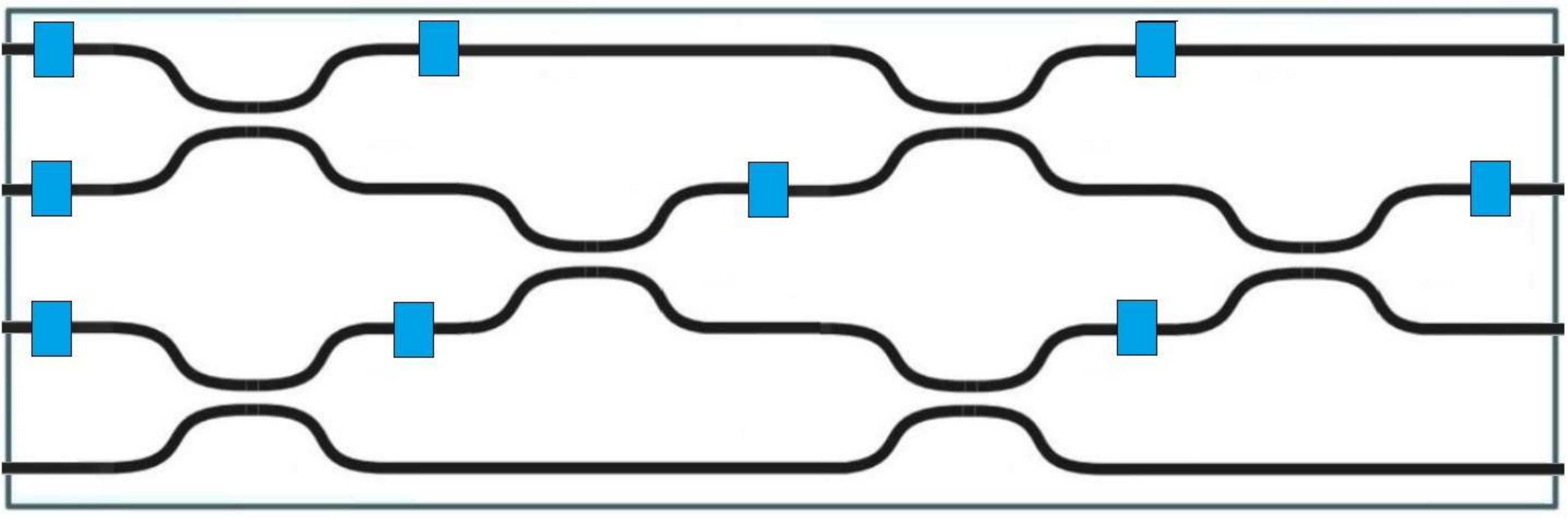}}
		\subfigure[ref3][$N=5$]{\includegraphics[width=8.5cm]{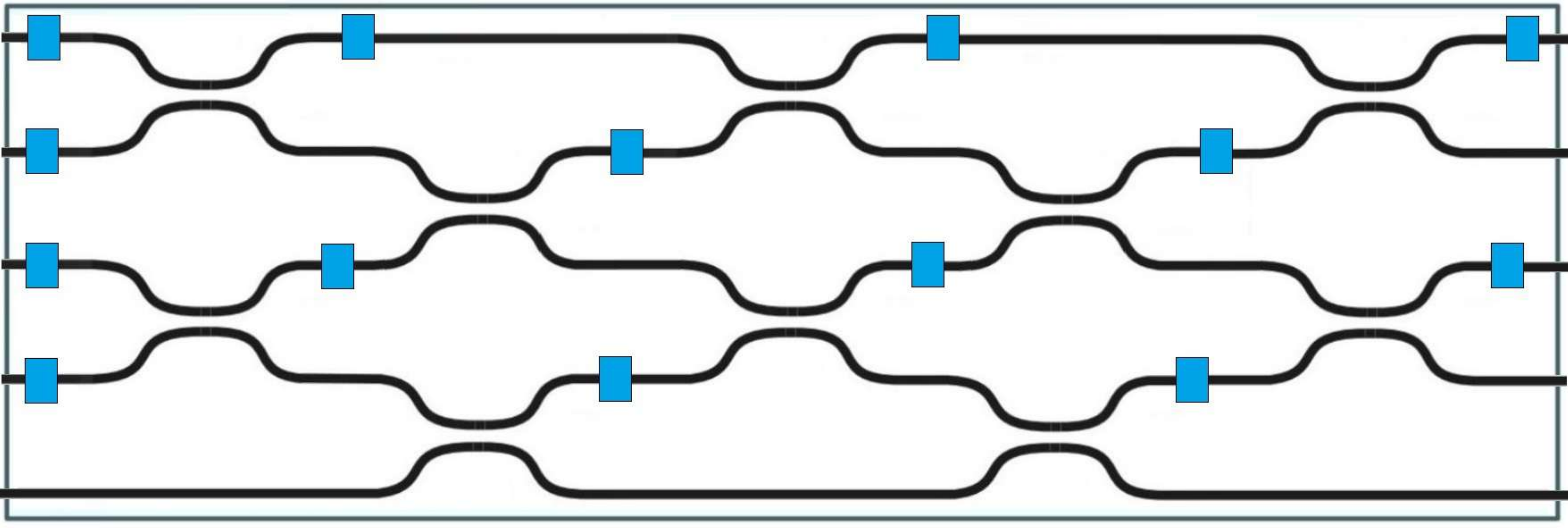}}
		\caption{General arrangement of beam splitters for realizing photonic circuits that implement Fourier transform for $N$-level quantum systems. The blue rectangles are phase shifters.}
		\label{fourier}	
	\end{figure}
	
	\begin{figure}[]
		\centering
		\includegraphics[height=2.8cm]{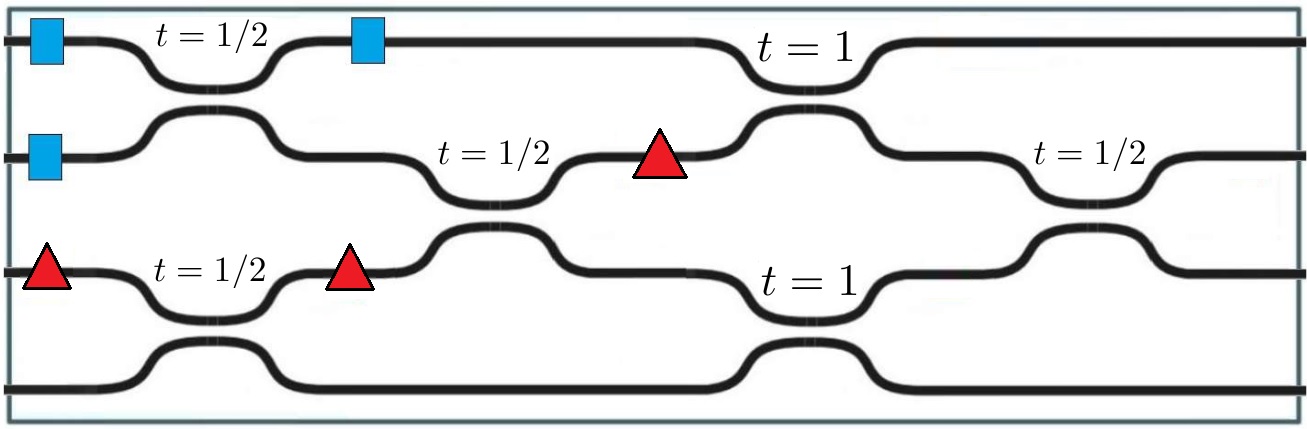}
		\caption{Specific beam splitters arrangement that implements the Fourier transform for a ququart system. The blue rectangles are phase shifters $\phi=\pi/2$ and the red triangles are phase shifters $\phi'=\pi$.}
		\label{quart}
	\end{figure}
	
	\begin{figure}[]
		\centering
		\includegraphics[height=3.15cm]{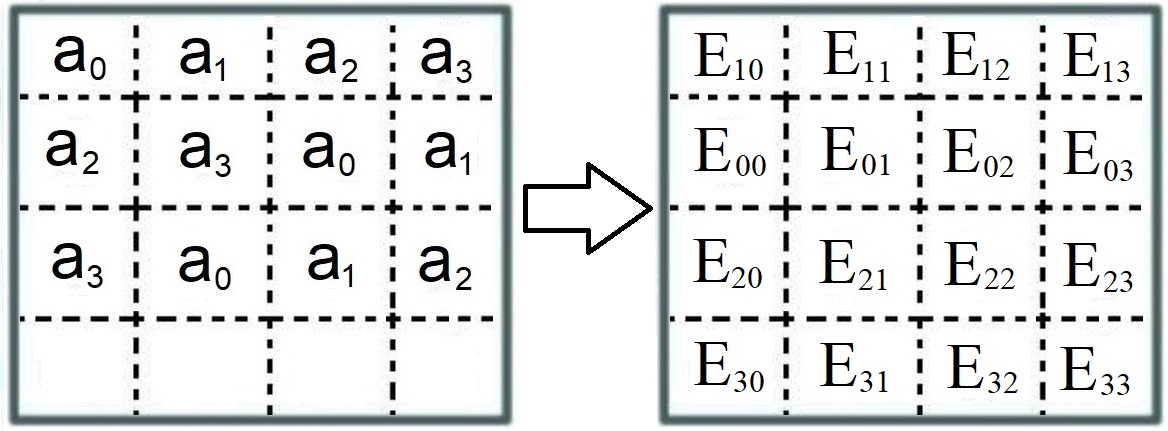}
		\caption{Schematic representation of the Fourier transform operation. The right frame shows the implemented POVM elements at each waveguide output. Single photon counts at this outputs are proportional to the photon detection probability in each interferometer output, i.e, are proportional to $\text{Tr}(\hat{\rho}\hat{E}_{ml})$, where the operator $\hat{E}_{ml}$ is defined in Eq.~\eqref{eij}. The proportionality constant, in this case, is equal to the sum of the single photon counts of all interferometer outputs.}
		\label{u3}
	\end{figure}	
	
	\begin{figure*}[]	
		\center
		\subfigure[ref1][Perspective view.]{\includegraphics[width=18cm]{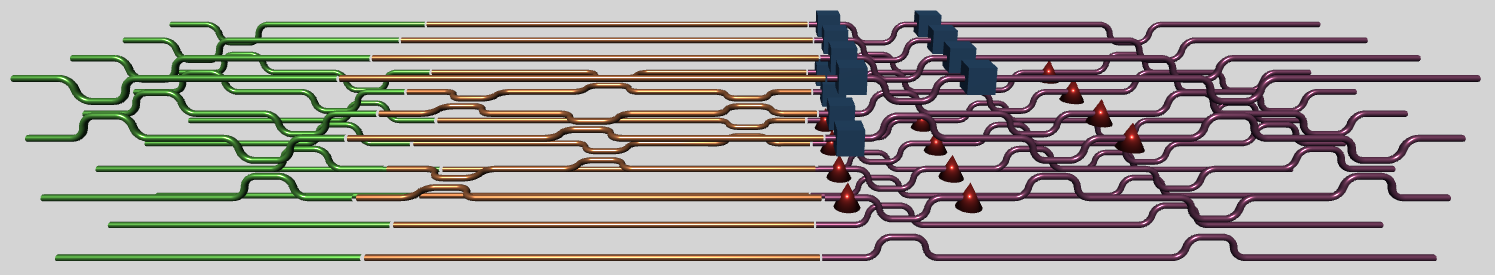}}
		\subfigure[ref2][Side view.]{\includegraphics[width=18cm]{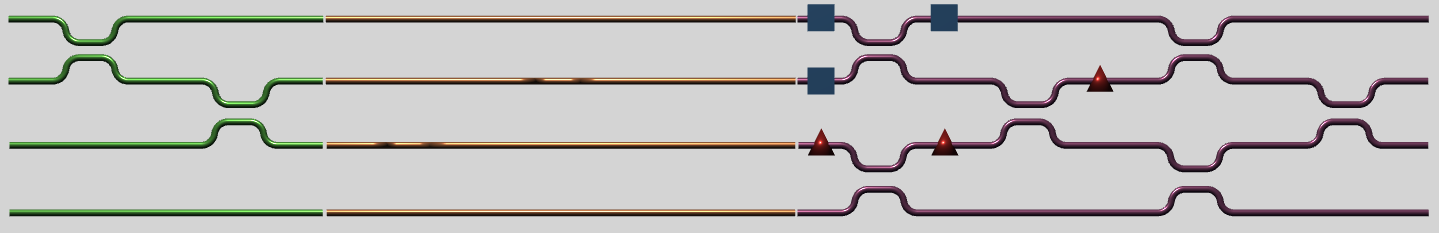}}
		\subfigure[ref3][Top view.]{\includegraphics[width=18cm]{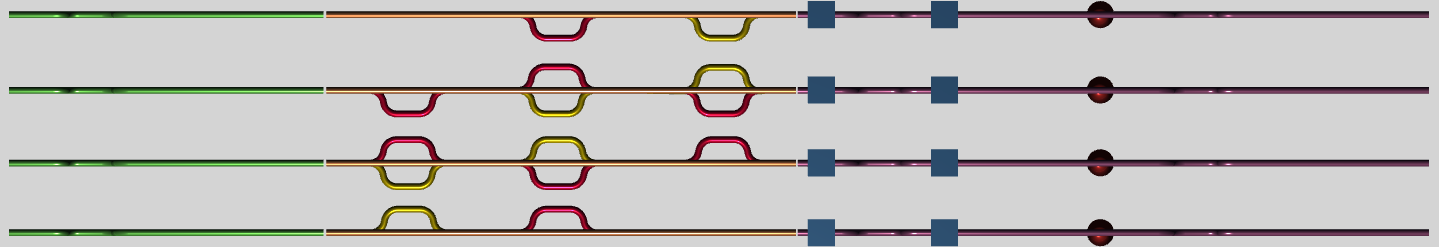}}
		\caption{Tri-dimensional scheme of a proposed photonic circuit for performing quantum state tomography of path ququarts system. The first part (decomposition) is presented in green, the second part (permutation) in orange and the third part (Fourier transform), in purple. The blue cubes represent phases shifters $\phi=\pi/2$ and red triangles, phases shifters $\phi'=\pi$. In (c), we adopted different colors to distinguish the waveguides of different layers in the second part of the circuit: the red guides belong to the layer $m=1$ and the yellow guides, to the layer $m=2$.}
		\label{3dququart}	
	\end{figure*}	

	Clements's method consists of realizing photonic circuits formed by combination of diagonals arrangements of beam splitters, so that the relation between the matrices representing these optical elements and the unitary operation $\hat{U}$ to be implemented is
	\begin{equation}
		\left(\prod_{\eta}\hat{T}_{\eta}\right)\hat{U}\left(\prod_{\eta'}\hat{T}_{\eta'}^{-1}\right)=\hat{D},
	\end{equation}
	where $\hat{D}$ is a diagonal matrix. The beam splitters represented by the matrices $\hat{T}_{\eta'}^{-1}$ are positioned from top to bottom, from left to right. On the other hand, the beam splitters represented by the matrices $\hat{T}_{\eta}$ are positioned from the bottom to top, from right to left. This circuit design is loss-tolerant and maintains high fidelity even for high dimensional quantum systems \cite{clements2016optimal}.
	
	Figure \ref{fourier} shows the arrangement of beam splitters and phase shifters that implement the last part of the circuit (for $N=3,4,5$), whose matrix representation is given by Eq.~\eqref{um}. As these arrangements are repeated in all vertical sectors, it has, in this last part, $N^2(N-1)/2$ beam splitters and the optical depth is $N$. 
	
	We continue with the ququart system analysis. The $\hat{U}$ matrix in Eq.~\eqref{um} for this case is 	
	\begin{equation}
	\hat{U}_4=\dfrac{1}{2}\begin{pmatrix}
	-i &  i & -1 & 1 \\
	 1 &  1 &  1 & 1 \\
	-1 & -1 &  1 & 1 \\
	 i & -i & -1 & 1 
	\end{pmatrix},
	\label{ft4}
	\end{equation}
	and the specific arrangement of beam splitters that implements it is presented in Fig.~\ref{quart}. Note that a permutation has been made between the first two rows of the matrix in Eq.~\eqref{um}. The acting of this operation is represented schematically in Fig.~\ref{u3}. In Fig.~\ref{3dququart}, it is presented a tri-dimensional scheme of a whole photonic circuit for performing quantum state tomography of path ququarts.
	
  	\subsection{General Advantages}
  	
  	\begin{figure}[]	
  		\center
  		\subfigure[ref1][]{\includegraphics[width=8.5cm]{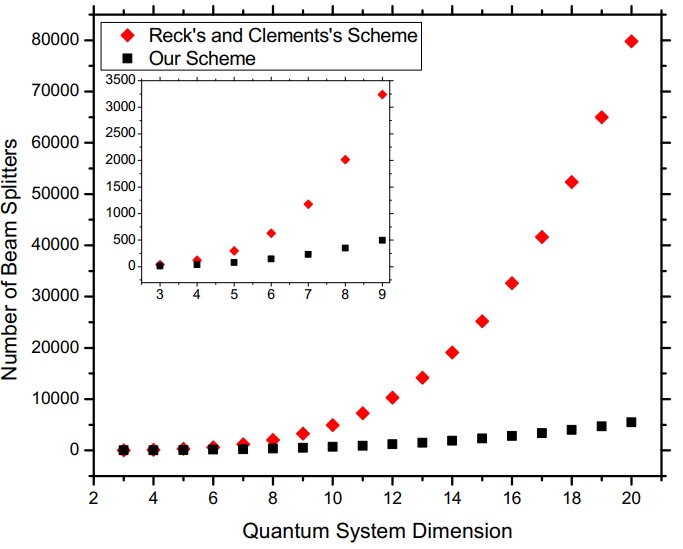}}
  		\subfigure[ref2][]{\includegraphics[width=8.5cm]{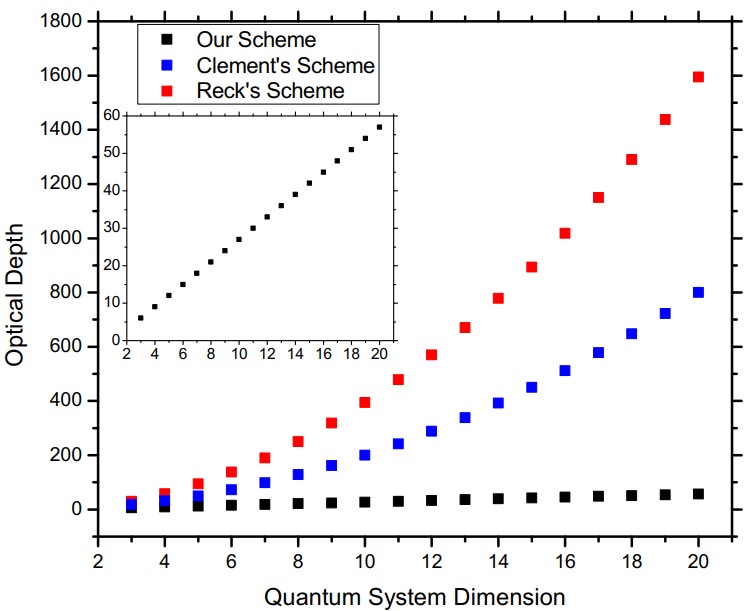}}
  		\caption{Comparison between the (a) number of beam splitters and (b) optical depth of the photonic circuit obtained by our 3D scheme and by the Clements's and Reck's schemes \cite{reck1994experimental,clements2016optimal}. The axes of the inset graphs shows, with more details, (a) the comparison between the number of beam splitters in lower dimensions and (b) our scheme optical depth.}
  		\label{compara}	
  	\end{figure}	
	
	With the information presented in the previous subsections, we can get an expression that gives the number of beam splitters $n_{\text{BS}}$ and the optical depth OD of an $N$-dimensional interferometer ($N\geqslant 3$). Such expressions are, respectively
	 \begin{equation}
	 n_{\text{BS}}(N)=\dfrac{1}{2}\left[N^3+N^2-4N+\sum_{\upsilon=1}^{N-2}2\upsilon(N-\upsilon)\right]
	 \label{nbs}
	 \end{equation}
	 and
	 \begin{equation}
	 \text{OD}(N)=3(N-1).
	 \label{po}
	 \end{equation}
	
	In order to demonstrate the main advantage of our proposal to generate more compact circuits, we compare the number of beam splitters and the optical depth of our scheme with those by Reck \textit{et al.} \cite{reck1994experimental} and Clements \textit{et al.} \cite{clements2016optimal}. In Fig.~\ref{compara}(a) is shown the comparison between the number of beam splitters required in each scheme. It is possible to notice that even for small values of $N$, our scheme presents itself as more economic than the others. This is due to the fact that the number of beam splitters scales as a degree 3 polynomial function with the quantum system dimension [Eq.~\eqref{nbs}], while, in the other schemes, a degree 4 polynomial function is obtained.	
	
	A comparison between the optical depths of the photonic circuits was also made and is presented in Fig.~\ref{compara}(b). While our scheme optical depth is given by a linear function in terms of $N$ [Eq.~\eqref{po}], the other schemes have a optical depth that changes quadratically as function of $N$. This corroborates the fact that our tomographic circuit is more compact than circuits conceived via other proposals. As a result, it implies that this photonic circuit has less losses than the others schemes for the realization of the quantum state tomography of qudits.
	
	\subsection{Errors and Losses}
	
	It is common knowledge that, in experiments with photonic circuits, the occurrence of errors and losses is commonplace and impossible to avoid. Thus, it is interesting to obtain methods for the determination of interferometers that guarantee a reduction in such occurrences. 
	
	To simplify the discussion of losses in photonic circuits, we will define two different types of losses: balanced and unbalanced losses \cite{clements2016optimal}. Balanced losses are those where all circuit paths experience equal losses. Thus, a large proportion of the balanced losses can be considered to be characterized by propagation losses. As such losses occur as a function of the circuit size, a proportionality relationship between propagation losses and the optical depth of the photonic circuit can be established. Therefore, we can conclude that, as shown in Fig.~\ref{compara}(b), the circuits designed using our method show a considerable reduction in their balanced losses, since their optical depth assumes very small values.
	
	On the other hand, when there are different losses in each circuit path, we say that these losses are unbalanced. These losses are typically caused by beam splitters, mainly due to bending losses and scattering \cite{clements2016optimal}. Thus, to evaluate the impact of unbalanced losses on the results of experiments with photonic circuits designed using our method, a simple model where all beam splitters experience equal losses was used. We calculate the fidelity of the transformation performed by the interferometer with imperfect beam splitters, described by the operation $\hat{U}_d$, in relation to the ideal interferometer, described by $\hat{U}$. The fidelity indicates how accurate the result of an experiment carried out with the lossy interferometer is in relation to the ideal case. For this calculation, we use the following expression	
	\begin{equation}
		F(\hat{U},\hat{U}_d)=\left|\dfrac{\text{Tr}\left(\hat{U}^\dagger\hat{U}_d\right)}{\sqrt{N\text{Tr}\left(\hat{U}_d^\dagger\hat{U}_d\right)}}\right|^2.
	\end{equation}
	
\begin{figure}[H]
		\centering
		\includegraphics[height=6.8cm]{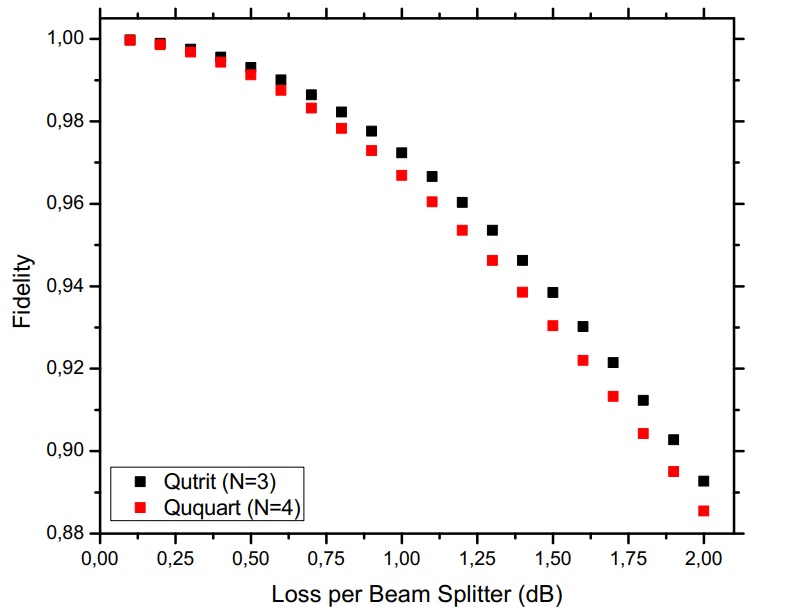}
		\caption{Fidelity as a function of loss per beam splitter in three-dimensional tomographical circuits for qutrit and ququart systems.}
		\label{fidelity}
	\end{figure}

	Figure \ref{fidelity} shows the relation between the fidelity $F(\hat{U},\hat{U}_d)$ and the loss per beam splitter, considered the same in all the beam splitters, for qutrit and ququart systems. Besides the fidelity in our photonic circuits are similar to that presented in the Ref. \cite{clements2016optimal}, it is noteworthy that, even in a case where each beam splitter experiences a loss equal to 2 dB, interferometers designed via the method presented in this paper have a fidelity of approximately 0.9.
	
	In addition to the photonic circuit intrinsic losses discussed in the previous paragraphs, errors inherent in the physical implementation are also present. To deal with such setbacks, an experimental test of this three-dimensional photonic circuit should be realized with single photon sources. Ideal single photon sources, i.e., a source that emit single photons on demand are still not available. Nevertheless, a two-photon light source produced, for example by a spontaneous parametric down conversion (SPDC), could be used. One of the photons will be transmitted by the photonic circuit and detected in coincidence with the second photon of the pair, that propagates in free space or in a fiber and its detection is used as a trigger what guarantees that the quantum tomography is done in the single-photon regime.
	
	\section{Conclusion}
	\label{sec:conclusao}
	
	In this work we were able to present a proposal for realizing quantum state tomography in 3D photonic circuits. We show that adopting the strategy of dividing the circuit into three different parts with operations acting independently in each vertical sector or layer, considerably reduces the complexity of the tomographic photonic circuit, as shown in Fig.~\ref{compara}.
	
	With the reduction of the photonic circuit complexity, one also reduces the difficulty of its production and the incurrence of losses and noise. In our scheme, the detection of the photons that emerge from each output of the circuit are made simultaneously, which optimizes the time spent in the experiment, also cooperating to reduce noise in the experimental data. This set of advantages shows the good applicability of our proposal in the design of 3D integrated circuits for performing quantum state tomography in path qudits.

	
	\begin{acknowledgements} This research was supported by the Brazilian agencies CNPq - Conselho Nacional de Desenvolvimento Cient\'{\i}fico e Tecnol\'ogico, Capes - Coordena\c{c}\~ao de Aperfei\c{c}oamento de Pessoal de N\'{\i}vel Superior, Fapemig - Funda\c{c}\~ao de Amparo \`a Pesquisa do Estado de Minas Gerais and INCT-IQ - Instituto Nacional de Ci\^encia e Tecnologia de Informa\c{c}\~ao Qu\^antica.
	\end{acknowledgements}
	
    
\bibliographystyle{unsrt}
\bibliography{Referencias}
\addcontentsline{toc}{chapter}{Referencias}
	
\end{document}